# A framework for the simulation of CR-VANET channel sensing, coordination and allocation

Hameer Abbasi[1], Shahid H. Abbassi[2], Ijaz M. Qureshi[2]

[1]Electrical Engineering Department, Military College of Signals, National University of Sciences and Technology, Adyala Road, Postal Code 46000, Rawalpindi, Pakistan

[2]Electrical Engineering Department, Air University, E-9, Postal Code 44000, Islamabad, Pakistan

## 1 ABSTRACT

VANETs are considered as networks of critical future importance, with the main concern being the safety of travelers and infrastructure. A lot of methods for coordination and channel allocation in the context of VANETs are being introduced. As such, the need of a framework to reliably compare the relative performances of different channel sensing, allocation and coordination schemes which takes into account the movement of vehicles is felt. In this paper, we introduce a framework that can be used to define and compare such schemes in a variety of scenarios. Simulation results clearly show the robustness of our technique by eliminating the misdetections and reducing to a great extent the false alarms.

## 2 KEYWORDS

VANETs, CR, Allocation, Sensing, Coordination, PU, SU

## 3 INTRODUCTION

Recent years have seen a rise in research in Vehicular Ad-hoc Networks (VANETs) for safety [1] and entertainment [2] applications. Current test VANET applications use the Dedicated Short Range Communications (DSRC) standard. [3] Since VANETs are inherently mobile wireless networks to be used in a consumer environment, the need for a bandwidth-efficient system is immediately felt. For this reason, there has been a lot of research into the possibility of using Cognitive Radios (CR) for effective spectrum utilization. A VANET environment where CR are used is termed as CR-VANET. As such, many different schemes of utilization have been proposed, as discussed in the section titled "Related Work".

---

[1] Correspondence: hameerabbasi@yahoo.com

There may be a gap between the methods used to produce results for such techniques, which may cause discrepancy in the data. For example, one person may use a different fading model to the next. Another may include small-scale fading while the first may not, which introduces additional discrepancy in the results. One may use a different mobility model to the next. One may use a different channel sensing technique to another. All of them are equally valid, but they lack a common platform for impartial comparison, which takes into account variables such as movement, fading, noise and other such factors which may cause variability in the simulation results; and at the same time allows for varying parameters, in order to perform simulations in a variety of conditions.

The historic approach in the scientific method has always been that the data being presented should be falsifiable. [4] These differences introduce a difficulty as to the falsifiability of the data, and even more importantly, the usefulness of such techniques as comparison is obscured.

## 4   RELATED WORK

Many researchers have put lot of efforts to analyse the sensing techniques used and fading models used for CR networks. There are a few works related specifically to use of CR in VANETs scenarios. Some existing works are discussed here.

In [5] it is proposed to use energy detection technique over a Gamma-shadowed Nakagami-composite fading channel for both large and small scale fading in VANETs. Authors have compared the results obtained in simulation with other composite fading models like Suzuki, Loo and Rice-Lognormal. In this paper, the authors have not discussed the coordination among vehicles for sensing the CR network. [6] proposes a model based on swarm-intelligence (the division of labour model in ant colonies), and claims to reduce congestion problems to the spectrum database. This paper proposes the use of cellular networks to access the spectrum database and to use IEEE 802.11p channels for inter-vehicular communication for cooperative sensing. Vehicles, after sensing, send the results to their neighbours. [7] pinpoints the drawbacks of the energy detection technique and proposes the use of the covariance based technique. According to the authors, the energy detection technique is sensitive to noise uncertainty. [8][9][10][8-10] Furthermore as signals are correlated in most practical applications hence energy detection should not be considered the method of choice. The authors propose the computation of the sample covariance matrix of the received signal based on the received signal samples; then to extract two test statistics from the sample covariance matrix. Finally, a decision is made for the presence or absence of the signal is made by comparing the ratio of two test statistics with a threshold.

In [11] it is proposed to assign a specific channel to each vehicle in TV spectrum band to each vehicle and sensing of the spectrum is performed independently by each vehicle. Information collected by each vehicle is exchanged among other neighbour vehicles to help with decisions on the future use of the spectrum. In [12] independent sensing by vehicles of Wi-Fi channels and ISM bands (2.4 or 5 GHz) has been proposed. Similar to [12], this paper also proposes the sharing of information among vehicles for the future use of spectrum holes. In [13] it is proposed that each vehicle sends a belief message about the presence or absence of a primary user (PU) signal to its neighbours. Each vehicle decides on the basis of its own observation and the belief messages received by it about the spectrum availability. Also, in [14] it is proposed that each secondary user (SU) senses and uses the spectrum holes on its own and shares the information with other SUs for future decisions. In [15] another independent spectrum sensing scheme based on the three state model is discussed. First state is empty slot or hole, second is presence of PU and the third one specifying presence of SU. Proposals discussed in [11-15] are all based on independent sensing of the PU spectrum. Since there is no coordination among the users, there is danger of sensing and occupying a single slot by more than one SU at a time causing harmful interference. These proposals may also create misdetections and create interference with licensed PU signals. The proposal discussed in [13] may suffer from slow convergence due to the broadcast of messages.

In [16] a cluster based scheme with one cluster head has been proposed. Every vehicle senses the spectrum and sends the result to its cluster head. The cluster head, based on results received from all the cluster members, assigns the hole to the requesting vehicle. This proposal is centralized and a hole detected by one vehicle may not be a hole for another vehicle due to different fading conditions at different locations. In [17] a spectrum division based scheme is proposed in which wide band spectrum is divided into several narrow sub-bands and a group of vehicles is assigned a sub-band to sense and use the spectrum. In this scheme a sub-band may suit a group of vehicles but may not suit others. In [18], a cooperation based scheme is proposed in which every vehicle senses the spectrum and sends the information to its neighbours. Each sample collected is assigned a weight which is adjusted and normalised. Based on the weight and output sensed decision is taken if the slot is empty or occupied and decision is forwarded to all the neighbours. This scheme may suffer from slow convergence and also lacks coordination. In [19], a scheme based on multi radio technologies has been proposed. Authors propose the usage different technologies for different classes of service, but in this paper no coordination or cooperation mechanism is discussed. In [20], a combination of standalone and cooperative sensing proposal have been studied. In this scheme a coordinator senses the spectrum and assigns a group of channels to the requesting vehicle. The requesting vehicle, on receipt of group of empty channels, re-scans the channels and uses the first available channel. This scheme suffers from the demerit that a hole for the coordinator may or may not be a hole for other member vehicles.

In this paper, we propose such a framework which varies the sensing and coordination method while keeping other factors constant. It will also allow simulations to be performed in a variety of scenarios, thus helping determine the robustness of a given technique. Our proposed scheme takes the localized decisions by employing three coordinators in a cluster; Main coordinator, forward coordinator and the backward coordinator. The simulation performed uses the Hata model for large scale fading and Rayleigh fading model for small scale fading. For mobility, we use the Gipps mobility model with a slight modification. In our proposal any amount of data may be recorded during a simulation, as desired. In summary, the framework will have the ability to:

- Perform simulations related to CR-VANET scenarios.
- Vary a number of factors, including, but not limited to:
    - The choice of mobility model.
    - The choice of coordination model.
    - The PU/SU scheduling algorithm used.
    - The choice of channel sensing algorithm
    - The channel allocation scheme

A rudimentary form of this simulator was used to produce the results in [21]. A version of those results using different scenarios, but with no change in the sensing or coordination algorithms, is produced in this paper.

# 5  THE PROPOSED FRAMEWORK

We first lay out a consistent framework for simulating different scenarios. In this framework, we will vary one factor and keep others constant to achieve consistency in the comparisons. We start with a "bird's eye" view of the entire algorithm, and then examine each process in finer detail, until we are down to the bare essentials. We do this by describing the mobility, sensing, coordination and scheduling algorithms individually, as well as when they are needed and what ties them in with the rest of the process.

The proposed framework starts as any time-based simulation model should, by dividing the entire time over which the simulation is to be performed into a number of time slices, and simulating over each of those slices.

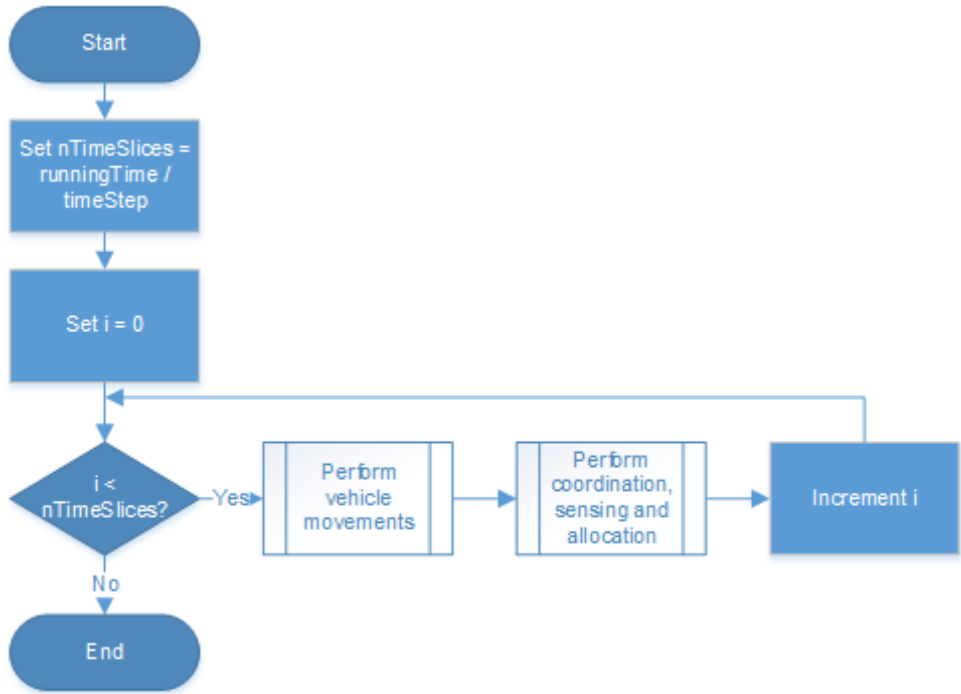

*Figure 1: Overall simulation framework.*

The psuedocode corresponding to the flowchart in **Error! Reference source not found.** is given below.

```
nTimeSlices := runningTime/timeStep
for i := 0 upto nTimeSlices - 1
    [Perform vehicle movements]
    [Perform sensing/coordination/allocation]
```

next

The procedure described in **Error! Reference source not found.** begins by calculating the number of time steps over which the simulation is to be performed, and then running repetitive procedures over each time slice. First, we move the vehicles as desired, and as dictated by the mobility model of choice. Second, we perform channel sensing, coordination and allocation as required by the different models used.

From first glance, the flexibility of this approach is immediately visible. We can choose the mobility model while performing movements, and we can also choose what methods to use for the other tasks. From here, the first sub-process is completely defined by the choice of mobility model to be used. The second one needs to be expanded, and it is shown below.

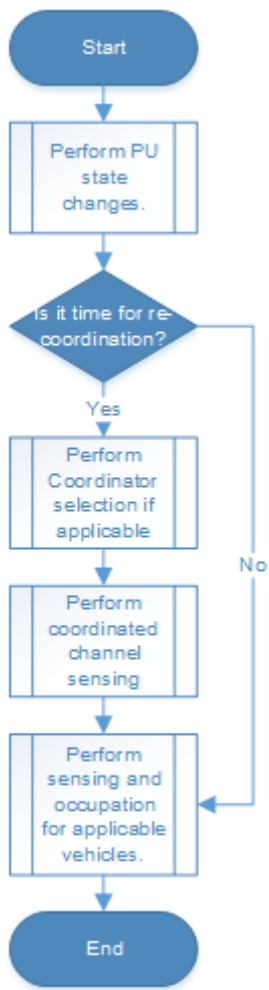

*Figure 2: Coordination and state changes.*

The psuedocode corresponding to the flowchart in **Error! Reference source not found.** is given below.

```
[Perform PU state changes]
if i*timeStep >= nextCoordinationTime
    [Perform coordinator selection]
    [Perform coordinated sensing]
end if
[Perform sensing and occupation for applicable vehicles]
```

The process shown in **Error! Reference source not found.** begins by performing the state changes for PU. This is defined as the PU (licensed) occupying or vacating the channel in the given time slice. After this, we check if coordination needs to be performed. If it does, we perform coordination and store results with the coordinators in a given range. We then perform sensing and occupation for individual vehicles that need to transmit over the CR spectrum.

Here, the coordinator selection sub-process is defined completely by the choice of coordination algorithm used. The other two need to be expanded, as they are below. Note that these processes are for individual channels or vehicles.

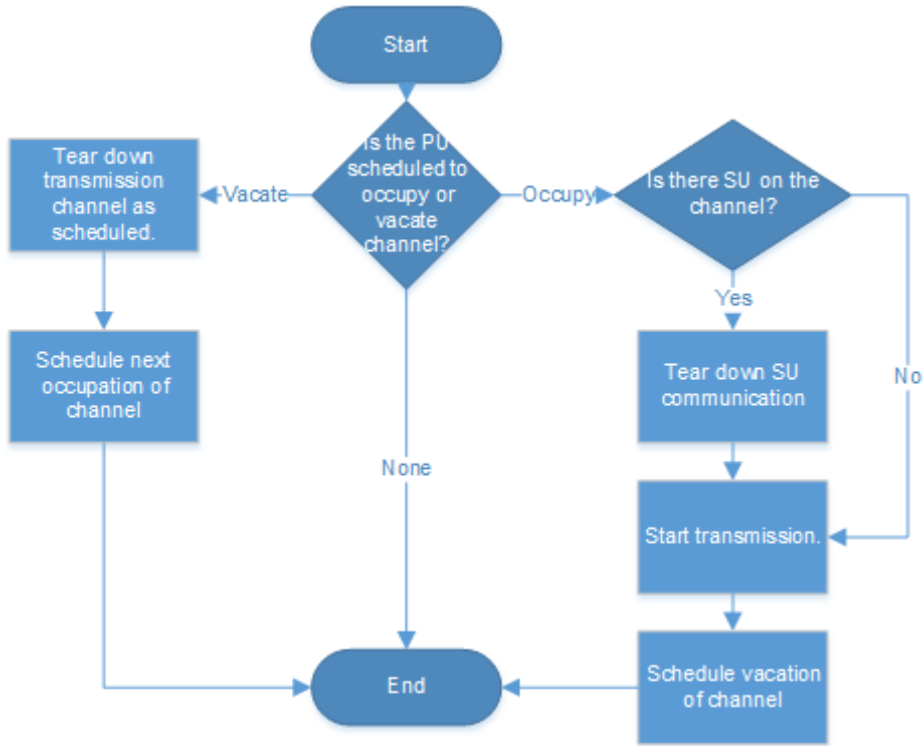

*Figure 3: Occupation and vacation*

The psuedocode corresponding to the flowchart in **Error! Reference source not found.** is given below.

```
if i*timeStep >= vacationTime && occupying

    [Tear down transmission channel]

    [Set occupationTime in accordance with scheduling algorithm]

    occupying := false

else if i*timeStep >= occupationTime && !occupying

    if occupied by SU

        [Tear down SU transmission]

    end if

    [Start transmission]

    [Set vacationTime in accordance with scheduling algorithm]

    occupying := true

end if
```

**Error! Reference source not found.** defines the basic scheduling algorithm used for PU state changes. It defines how the simulation will decide when a PU is to occupy a channel, and when it is to vacate a given channel.

Again, a lot of flexibility as to the method of scheduling transmissions is allowed. A message must be sent to SUs occupying this channel, possibly over the WAVE control channel.

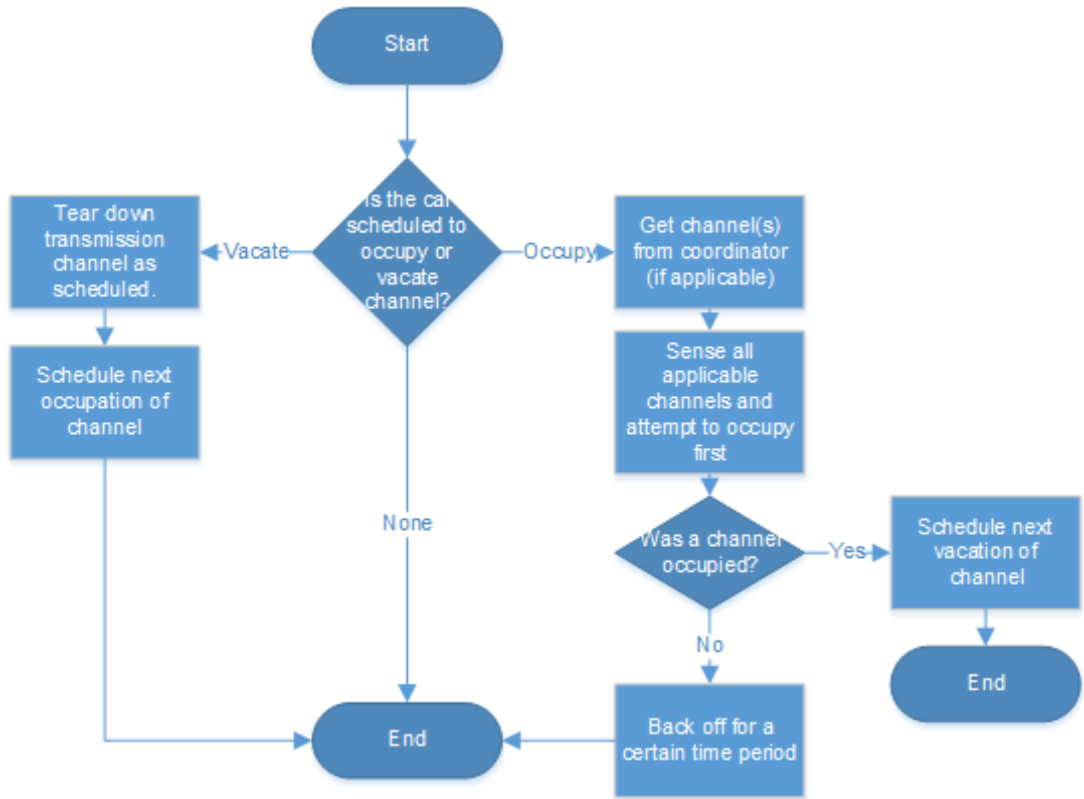

*Figure 4: Transmission and back off mechanism.*

The psuedocode corresponding to the flowchart in **Error! Reference source not found.** is given below.

```
if i*timeStep >= vacationTime && occupying
    [Tear down transmission channel]
    [Set occupationTime in accordance with scheduling algorithm]
    occupying := false
else if i*timeStep >= occupationTime && !occupying
    if coordinated sensing scheme
        [Get channel(s) from coordinator]
```

```
            end if
        [Attempt to sense/Occupy as necessary]
        if occupation was successful
                [Set vacationTime in accordance with scheduling
                    algorithm]
                occupying := true
        else
                [Extend occupationTime by back-off time]
        end if
    end if
```

These procedures, shown in **Error! Reference source not found.**, define the (more complicated) scheduling and allocation scheme for SUs of the CR spectrum. It takes into account the fact that a channel may not always be allocated to a SU when needed, and it also lays the groundwork for what sensing is to be performed before a SU will attempt to occupy a channel.

Please note that this last flowchart may vary slightly depending on the needs of the coordination and allocation scheme. For example, coordination may not be required at all, multiple channels may be sensed before a back-off is triggered, and so on. The flowchart has been made for the requirements of [21], but the pseudo-code is more generalized. This also takes into account the possibility of a misdetection.

## 6    MODELS USED FOR THE SIMULATION

The Hata Model for suburban areas was used as the large-scale fading model in this simulation. This depends on the Hata model for urban areas. This gives the predicted median path loss over a given path. [22]

Whereas the Hata model is a large-scale model that predicts averages over a given distance, Rayleigh fading predicts sudden changes due to rapidly varying channel conditions. [23]

Rayleigh fading is a long-standing model used in wireless communication systems for power prediction, where a line-of-sight component is absent. Considering the scenario, which is a suburban environment, it is highly unlikely that one will be present. Therefore, we have chosen the Rayleigh model for our simulation.

The Gipps mobility model describes how traffic behaves in a typical vehicular scenario. [24]

The Gipps model is one of the classical models used to simulate urban mobility. We choose it for its simplicity and effectiveness in predicting vehicle behaviour. We made one modification to this model for our purposes: That the vehicle will not go below a certain minimum speed threshold. Given we are simulating a highway scenario where vehicles may overtake each other, this assumption is reasonable. We propose Equation 1 as a replacement for $v(t)$ in our simulations.

$$v(t) = \max[v_{\min}, \text{rand}[v_{\text{des}}(t) - \epsilon a, v_{\text{des}}(t)]]. \qquad (1)$$

## 7 SIMULATION RESULTS

Here, we give simulation results for three different sensing and allocation schemes, for three different kinds of data. We vary the number of available channels, the number of vehicles in the simulation and the average speed of the vehicles. The schemes used are standalone sensing, cooperative sensing, and proposed sensing, as used in [21]. The other parameters were: Vehicle sensing range was 400 m and communication range was 240 m, the length of road used was 2 km, the noise power was -133.16 dBW (500 K system noise temperature at a bandwidth of 7 MHz). Two PU transmission towers were used, each 10 km away from the closest point on the road. The mobility model used was the Gipps model [24] and the wave propagation model used was the Hata model [22] for large scale fading and the Rayleigh model [23] for small scale fading. The frequency used was 150 MHz, the base station height was 50 m and the mobile user height was 1.5 m. The average speed for vehicles was set to 100 km/h ± 20% per vehicles (when not varied), the number of channels to 100 (50 per PU tower) and number of vehicles to 50 (25 per side). A vehicle could deviate to within 10% of its average speed. The reaction time was set to 1 second, and the deceleration function was 2.12 times velocity. The human error factor could vary between 0.25 and 0.4 for each vehicle. The vehicle back-off time was 10 ms and the coordination time was 20 ms.

The recorded data in the simulations was

a) The number of successful allocations of a channel.
b) The number of false alarms (a channel was vacant but was sensed as occupied).

c) The number of misdetections (a channel was occupied but was sensed as vacant). Misdetections are dangerous because they can interfere with PU or other SU communications.

The following figures 5 to 13 show the results of simulations run under a number of scenarios.

As one can clearly see, in all three cases, the number of mis-detections is virtually eliminated. In this limited simulation, they actually came out to be zero in all cases. This is shown in **Error! Reference source not found.**, **Error! Reference source not found.** and **Error! Reference source not found.**. The number of false alarms is also greatly reduced, as shown in **Error! Reference source not found.**, **Error! Reference source not found.** and **Error! Reference source not found.**. One can see a slight fall in the number of allocations as compared to standalone sensing in **Error! Reference source not found.**, **Error! Reference source not found.** and **Error! Reference source not found.**. This is because coordination may have an adverse effect on the allocations, due to the time taken to communicate channels to and from the coordinator. In any case, the benefits outweigh this slight disadvantage, particularly that of the elimination of misdetections. It is very important that a SU should not interfere with a PU's communication. This is achieved in the proposed sensing model.

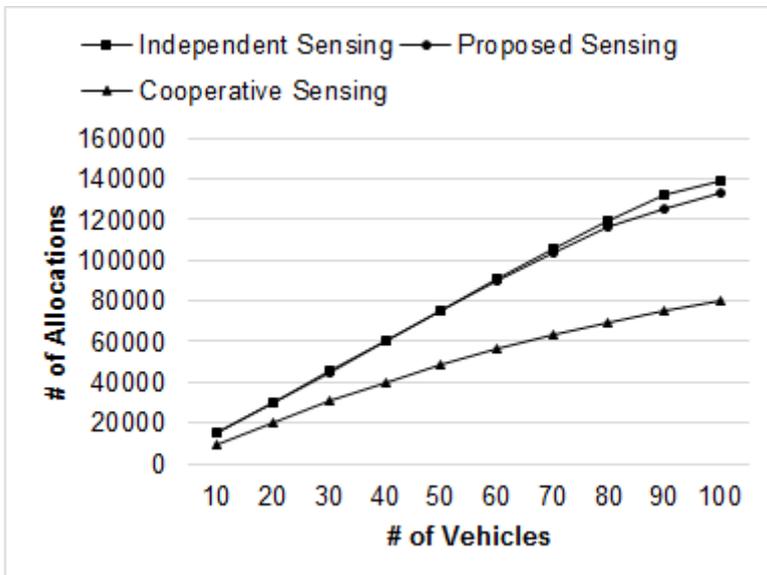

*Figure 5: Number of number of vehicles vs. number of allocations.*

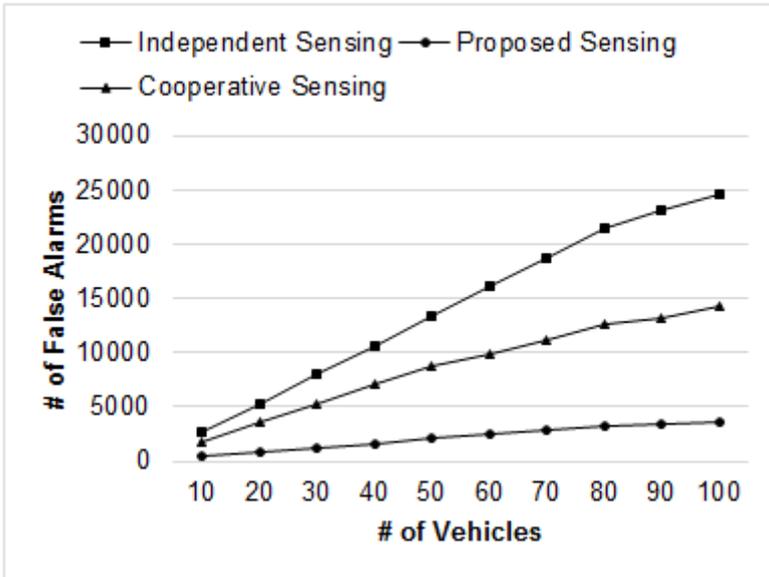

*Figure 6: Number of number of vehicles vs. number of false alarms.*

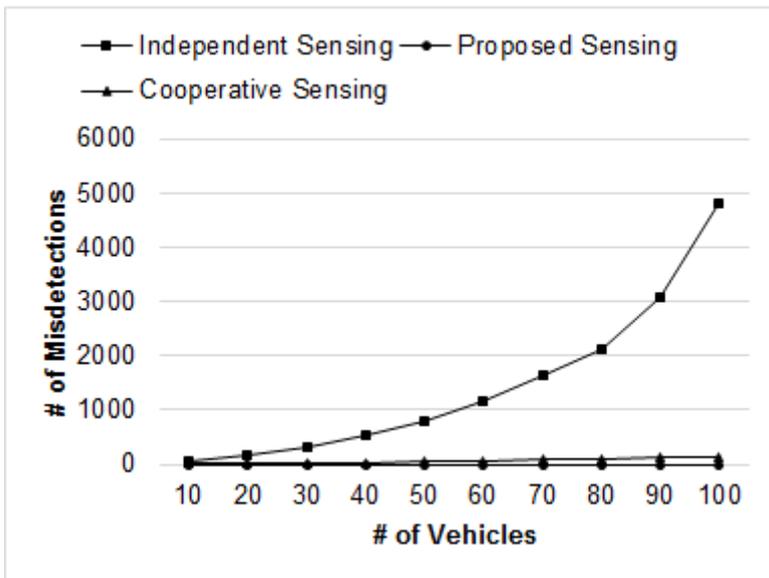

*Figure 7: Number of Number of number of vehicles vs. number of mis-detections.*
3

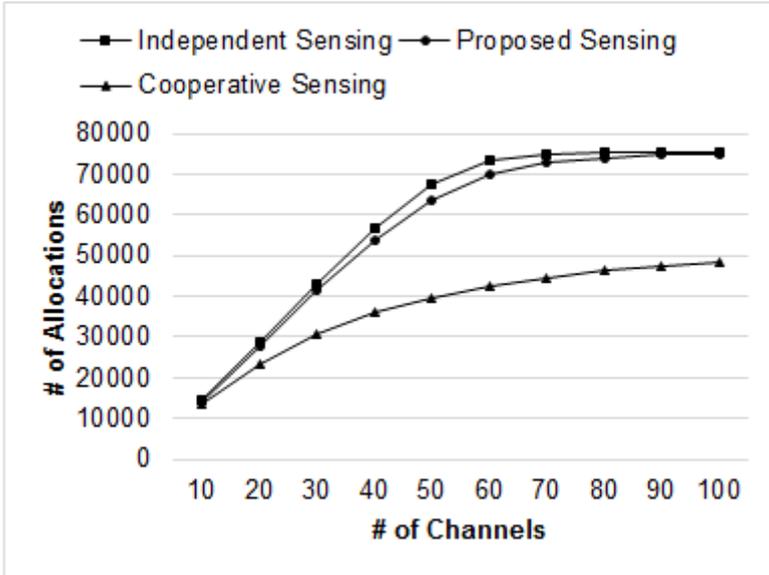

*Figure 8: Number of channels vs number of allocations.*

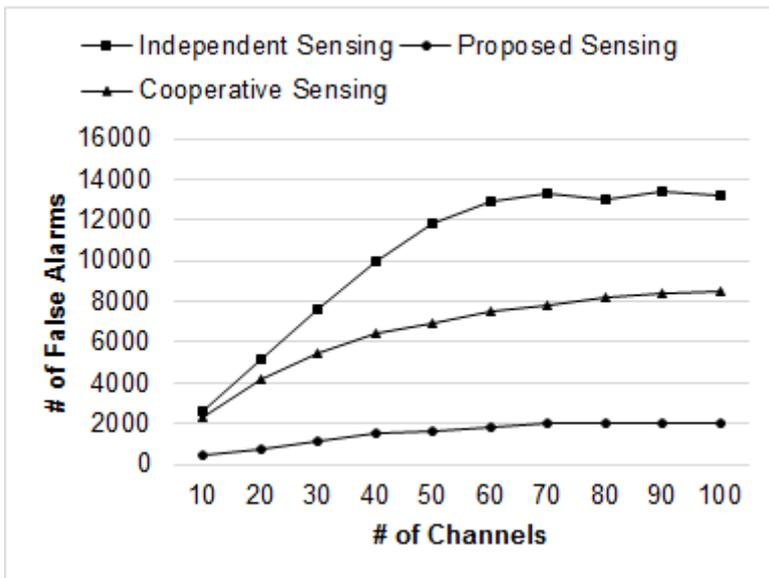

*Figure 9: Number of channels vs number of false alarms*

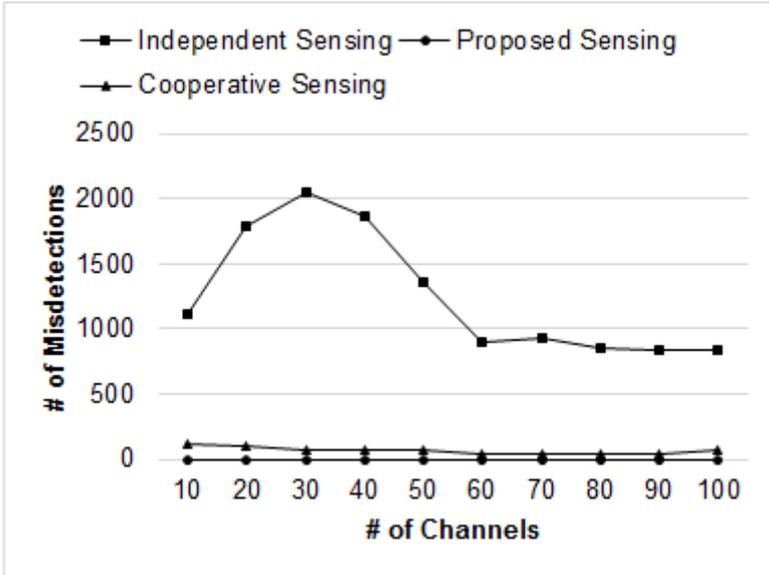

*Figure 10: Number of channels vs number of mis-detections.*

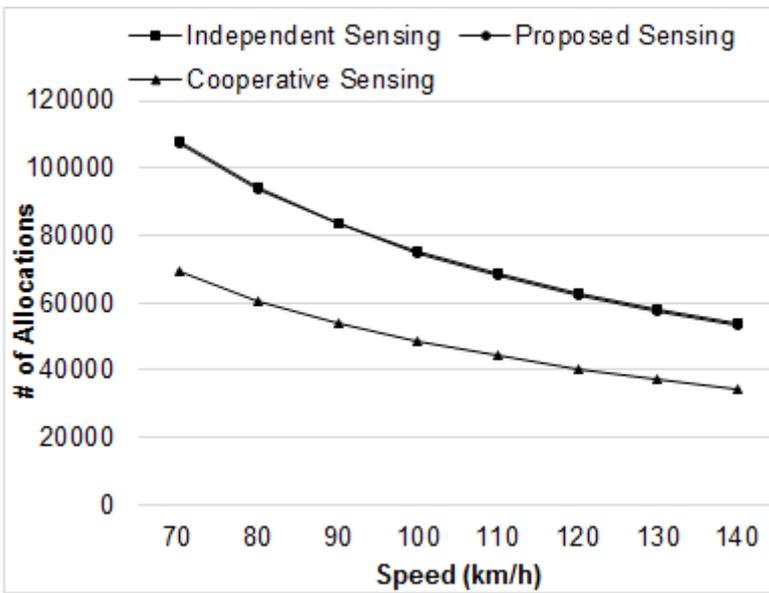

*Figure 11: Speed vs number of allocations.*

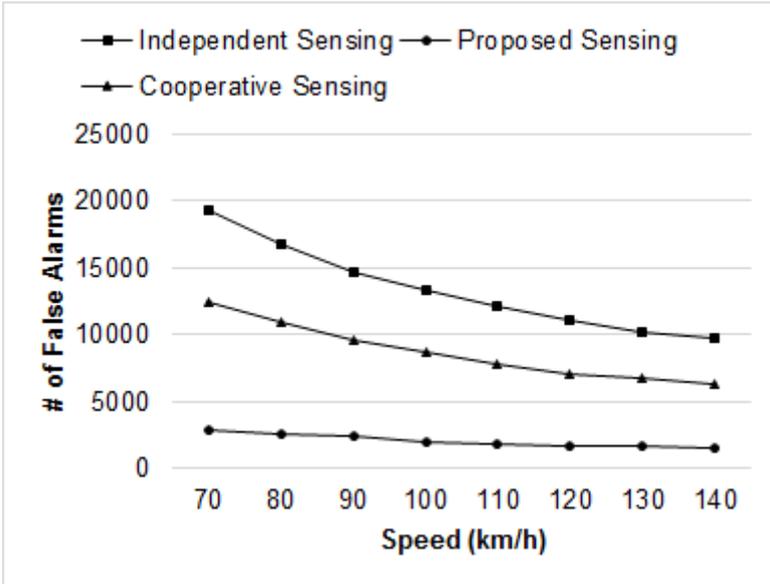

*Figure 12: Speed vs number of false alarms.*

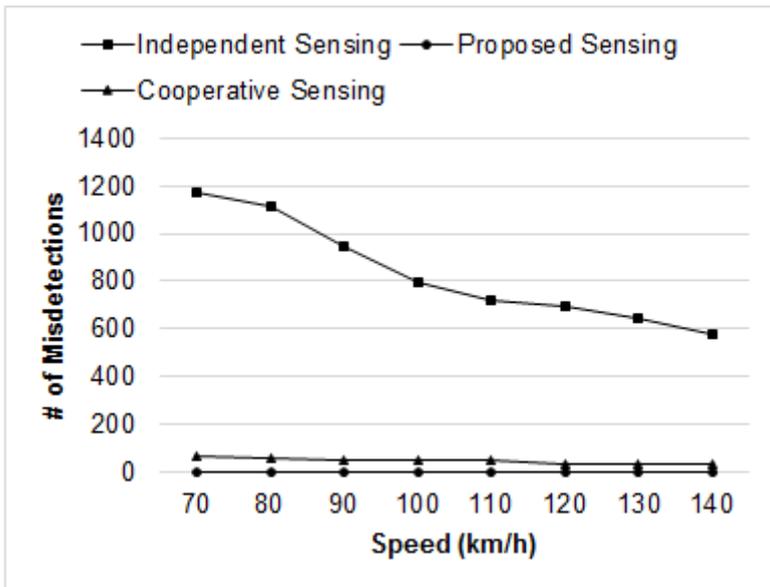

*Figure 13: Speed vs number of mis-detections.*

## 8  CONCLUSION

In this paper, we introduced a framework for the impartial comparison of CR-VANET sensing and allocation schemes. We described the framework and all the models that could be used to fill in various parts of the algorithm. We then described the schemes we used to produce a complete simulation. We then ran simulations in line with the framework for three different channel sensing and coordination schemes. The proposed coordination scheme with three coordinators per cluster showed marked benefits over independent and cooperative sensing schemes.

## 9 FUTURE PROSPECTS

As the concept of the application of CRs is relatively new to VANETs, there is a lot of room for improvement. Further work needs to be done on the application of effective spectrum management and allocation in VANETs. This is particularly true for CR-VANETs, as there is a possibility of PU involvement. Furthermore, work needs to be done into the possibility of using bandwidth-efficient techniques such as OFDM in CR-VANETs. As power generation isn't a huge issue, work needs to be done on immersive entertainment systems while inside a vehicle, while improving the safety. Vehicle to vehicle communication in real-time may open up the door for maintaining a safe distance from other vehicles, with facilities for auto-braking in the case of risks. Other similar features for lane guidance and optimal speed may also help improve the utility of vehicles in general.

*Transportation Research,* pp. 105-111, 1981.